# Observing Spin Polarization of Individual Magnetic Adatoms

Y. Yayon[1], V. W. Brar[1], L. Senapati[2], S. C. Erwin[2], and M. F. Crommie[1]

[1]Department of Physics, University of California at Berkeley, Berkeley, California 94720-7300

and Materials Sciences Division, Lawrence Berkeley Laboratory, Berkeley, California 94720-7300

[2]Center for Computational Materials Science

Naval Research Laboratory

Washington DC 20375-5000

We have used spin-polarized scanning tunneling spectroscopy to observe the spin-polarization state of individual Fe and Cr atoms adsorbed onto Co nanoislands. Both of these magnetic adatoms exhibit stationary out-of-plane spin-polarization due to their direct exchange interaction with the substrate, but the sign of the exchange coupling between electron states of the adatom and the surface state of the Co island is opposite for the two: Fe adatoms exhibit parallel spin-polarization to the Co surface state while Cr adatoms exhibit antiparallel spin-polarization. First-principles calculations predict ferromagnetic and antiferromagnetic alignment of the spin moment for individual Fe and Cr adatoms on a Co film, respectively, implying negative spin-polarization for Fe and Cr adatoms over the energy range of the Co surface state.

Since the early tunnel-junction measurements of Julliere (*1*), much work has been aimed at reducing the size of magnetic tunnel junctions (MTJ). This has led to new possibilities for performing fundamental explorations into the quantum spin behavior of atomic and molecular systems (including quantum information applications) (*2-4*) as well as new possibilities for improved performance of magnetic-field sensors, magnetic memory, and new spintronic devices (*5*). An important goal in this trend is the eventual creation of devices whose functionality can be engineered at the level of individual atomic spins (*5*). Measuring the spin polarization state of individual atoms and understanding how atomic spins behave in a condensed matter environment are essential steps toward this goal.

Much progress has already been achieved through atomically-resolved spin-polarized scanning tunneling microscopy (SP-STM) measurements of magnetic thin films (*6, 7*). Additionally, Spin-coupling of conduction electrons to single magnetic atoms has been observed via the Kondo effect (*8, 9*), and spin-flips have been observed via inelastic tunneling through individual magnetic atoms (*10, 11*) and magnetic resonance force microscopy of silicon defects (*12*). Direct observation of the spin-polarization state of isolated adatoms, however, remains challenging, in part because isolated atoms have low magnetic anisotropy energy (MAE) which causes their spin to fluctuate in time due to environmental interactions. Recent measurements of adatom MAE's range from less than 1 meV/atom to 9 meV/atom, but the techniques used in these measurements are unable to resolve the spin-polarization state of an individual magnetic adatom (*11, 13*).

Here we report a measurement of the spin-polarization state of individual Fe and Cr adatoms on a metal surface. In order to fix the adatom spin in time, the adatoms were deposited onto ferromagnetic (FM) Co nano-islands, thereby coupling the adatom spin to the island magnetization through the direct exchange interaction. Low temperature spin-polarized scanning tunneling spectroscopy (SP-STS) (*6*) was used to probe the local spin-dependent electronic structure of isolated Fe and Cr adatoms prepared in this way. Clear spin-polarized (SP) contrast is seen between the two possible spin states (up and down) of each magnetic species. The two types of magnetic atoms, however, show dI/dV spectra that differ greatly in their spin-averaged local density of states (LDOS) and spin-polarization contrast. Furthermore, the two types of atoms show opposite spin coupling to the underlying cobalt substrate. Over the energy range that includes the cobalt island SP surface state, Fe atoms display spin-polarization parallel to the cobalt substrate electrons while Cr atoms display antiparallel spin-polarization. First-principles calculations predict FM coupling between an Fe adatom and a Co island, and antiferromagnetic (AFM) coupling between a Cr adatom and a Co island, implying negative spin-polarization (minority LDOS is larger than majority LDOS) around -0.3 eV below the Fermi energy for Fe and Cr atoms.

Our experiments were conducted using a modified commercial low-temperature STM (OMICRON LT-STM). All measurements were performed in ultrahigh vacuum (UHV) ($<10^{-11}$ mb) and at low temperature (4.8 K). SP tips were created by coating etched tungsten tips with a thin film of Cr to produce an out-of-plane magnetization as described by O. Pietzsch, *et al.* (*14*). The Cu(111) single crystal was cleaned in UHV by cycles of Ar ion sputtering and thermal annealing. Magnetic islands were obtained by depositing Co onto Cu(111) at room temperature (*14-16*). Fe and Cr atoms were deposited by e-beam evaporation while the sample was kept cold (~10K). The differential conductance (dI/dV) signal was measured through lock-in detection of the ac tunneling current modulated by a 477 Hz, 5mV (rms) signal added to the junction bias (bias voltage here is defined as the sample potential referenced to the tip). I(V) and dI/dV spectra were measured by fixing the tip position at one point and scanning the junction voltage under open loop conditions. dI/dV images were acquired by spatially scanning the STM tip at constant current and measuring the dI/dV signal as a function of lateral position.

Cobalt islands were chosen as a substrate because their SP electronic structure has been well-studied theoretically and experimentally (*14, 16-18*), thus providing a calibrated substrate where different magnetization states ("up" and "down" with respect to the surface plane) are easily accessed. Fig. 1 shows a representative 36 nm x 36 nm topograph of a 0.002 monolayer coverage of magnetic adatoms (Fe in this case) adsorbed onto triangular Co islands on the Cu(111) surface. The magnetic adatoms can be seen protruding from the surface of the islands and the surrounding copper. Spatial oscillations seen on the Cu(111) surface are due to interference of surface-state electrons scattered from the adatoms and Co islands (*19, 20*).

The SP electronic structure of individual Fe and Cr adatoms on cobalt nanoislands was characterized using SP-STS as seen in Fig. 2. All dI/dV spectra were obtained from islands oriented spatially in the same crystallographic direction to avoid structure-induced contrast (i.e., due to fcc vs. hcp stacked islands) (*14, 16*). Spin-up and spin-down Co islands were distinguished spectroscopically via contrast arising from a spin polarized surface state centered 0.28 eV below the Fermi energy, as seen in Fig. 2A (dashed lines) (*14, 16*). This resonance is believed to arise from a SP minority surface state of $d_{3z^2-r^2}$ symmetry (*16*). SP-STS at high magnetic field showed that the magnetizations of the Co island and Cr tip are parallel when the dI/dV spectrum shows higher intensity around the island surface state and antiparallel for lower intensity, implying that both Co islands and Cr tips have negative magnetization at that energy range (*14*). Therefore, if we define the tip magnetization direction as down $(\downarrow)$ then the magnetization of the Co islands displaying a higher magnitude surface-state peak is $(\downarrow)$, corresponding to total island spin parallel to total tip spin, and the magnetization of islands displaying a lower magnitude peak is up $(\uparrow)$, corresponding to total island spin antiparallel to total tip spin. SP spectra measured for individual Fe adatoms (Fig. 2A, solid lines) show strong spin-polarization contrast depending on whether they lie on $\downarrow$ or $\uparrow$ Co islands. SP spectra measured for individual Cr adatoms on Co islands (Fig. 2B, solid lines) also show strong spin-polarization contrast depending on the magnetization of the island on which they lie. The magnetic contrast for Cr atoms however, qualitatively differs from the contrast seen for Fe atoms.

These spectroscopic differences in adatom behavior are best seen by normalizing the spectra according to the scheme of ref. (*21*) which is useful in predicting spin contrast in dI/dV maps from dI/dV point spectroscopy. This normalization scheme is based on the fact that the dI/dV spectrum changes between two points on the sample by the same factor as the I(V) spectrum and is given by the following formula:

$$\left(\frac{d\tilde{I}}{dV}\right)_i = \left(\frac{dI}{dV}\right)_i \frac{I_1(V)}{I_i(V)} \qquad \text{where } i = 1,2,....4. \qquad (1)$$

Here i indexes a specific spectrum and $\left(d\tilde{I}/dV\right)_i$ is the corresponding normalized spectrum. $I_i(V)$ is measured simultaneously with each corresponding $(dI/dV)_i$ spectrum, and the choice of $I_1(V)$ does not affect the ratio between normalized $dI/dV$ spectra (*21*). In the energy range where the Co island spin-polarization is most pronounced (-0.35 eV to -0.20 eV with respect to $E_F$) Fe adatoms exhibit spin-polarization in the same direction as the Co island spin contrast (as seen in Figs. 2A and 2C): Fe atoms on ↓ islands have a stronger dI/dV signal than Fe atoms on ↑ islands. Cr atoms, on the other hand, show opposite spin behavior as seen in Figs. 2B and 2D: Cr atoms on ↓ islands exhibit a lower dI/dV signal than Cr atoms on ↑ islands.

The reversed spin contrast behavior for Fe and Cr adatoms on Co islands is summarized in the difference spectra plotted in Fig. 2E. The difference between SP spectra measured on ↓ islands relative to ↑ islands is shown for both adatom species compared to the Co island difference spectrum. In the energy range of maximum Co island spin-contrast the Fe adatoms are seen to have the same sign of spin contrast as the Co islands while the Cr adatoms display spin contrast with an opposite sign. Over this energy range the Fe adatom spin polarization is thus parallel (FM coupled) to the Co island while the Cr adatom spin polarization is antiparallel (AFM coupled) to the Co island magnetization. Similar spin contrast was seen for hundreds of Fe and Cr atoms on more than twenty different islands using five different SP-tip preparations. SP-STS of Fe and Cr adatoms on the bare Cu(111) substrate showed no spin contrast and non-SP spectra (Fig. 2F) also did not show any discernible contrast for Fe and Cr adatoms on different Co islands.

The reversed SP behavior of individual Fe and Cr adatoms at the energy range of the Co island surface state can be clearly seen in spatially-resolved SP dI/dV maps. Fig. 3A shows a color-scaled SP dI/dV map together with topograph contour-lines (measured simultaneously) for Fe and Cr atoms co-deposited on two Co islands (the two Co islands have opposite magnetization orientation). Fe and Cr atoms can be easily distinguished by their topographic signatures (Cr atoms protrude 0.07 nm from the island surface while the Fe atoms protrude 0.04 nm). Spin contrast between adatoms sitting on the two oppositely polarized islands is seen in linecuts through Fe and Cr atoms shown in Fig. 3B-E. Fe atoms sitting on the ↓ island exhibit a larger dI/dV signal than Fe atoms on the ↑ island, while Cr atoms on the ↓ island show a smaller dI/dV signal than Cr atoms on the ↑ island. This further confirms the parallel nature of the Fe-adatom/Co-island SP and the antiparallel nature of the Cr-adatom/Co-island SP in this energy range. Thus we conclude that SP-STS clearly reveals single adatom spin contrast: each type of adatom reveals a distinct spectrum, and over the energy range of the Co island surface state Fe and Cr adatoms show opposite SP directions. However, this measurement does not unambiguously determine the direction of the total spin of the adatom, because the total spin is an integral over all filled states while the spectra shown here were taken over a limited energy range.

To theoretically investigate the spin-coupling of single Fe and Cr adatoms to a ferromagnetic 2-ML film of Co on Cu(111), we performed density-functional theory calculations within the generalized-gradient approximation using the projector-augmented wave method (*22-24*). Adsorption binding energies were calculated with adatom spin held both parallel and antiparallel to the magnetization direction of the Co film. In each case full relaxation was carried out for the adatom and the topmost three atomic layers, using 3x3 supercells and 2x2 sampling of the surface Brillouin zone. The resulting energies (shown in Fig. 4) show that Fe adatoms prefer ferromagnetic alignment to the Co film whereas Cr adatoms prefer antiferromagnetic alignment. For the calculations to be consistent with the SP measurements it implies that over the energy range of the Co island surface state Fe and Cr adatoms should exhibit negative magnetization. Within an effective Heisenberg spin-Hamiltonian the energy difference between spin-parallel (FM) and spin antiparallel (AFM) alignments is equal to twice the exchange coupling energy, J, between an adatom spin and the effective spin of the Co film. This gives J = -0.5 eV for Fe adatoms and J = 0.25 eV for Cr adatoms, favoring FM and AFM adatom spin alignment with the cobalt film magnetization, respectively. These high exchange energies explain how we are able to observe adatom SP despite our environmental temperature of 4.8 K.

In conclusion, we have used SP-STS to observe the spin-polarization state of individual Fe and Cr adatoms adsorbed onto Co nanoislands. Over the energy range of the Co island surface state, Fe adatoms exhibit SP parallel to the Co island while Cr adatoms exhibit antiparallel SP. Calculations predict FM and AFM alignment of the spin moment for Fe and Cr adatoms on a Co film, respectively, implying negative spin-polarization for Fe and Cr adatoms over the energy range of the Co surface state. The ability to measure the SP state of individual adatoms together with the capability of STM-based atomic manipulation (*25*) opens new possibilities for probing the magnetic properties of nanostructures constructed from individual atoms, and molecules.

This work was supported in part by NSF Grant EIA-0205641, by the Director, Office of Energy Research, Office of Basic Energy Science, Division of Material Sciences and Engineering, U.S. Department of Energy under contract No. DE-AC03-76SF0098, by the Office of Naval Research, and by the National Research Council. Computations were performed at the DoD Major Shared Resource Center at ASC.

**Figure legends:**

**Fig. 1:** Topograph of Fe adatoms adsorbed onto triangular Co islands on Cu(111) at T=4.8 K. Fe adatoms are seen as green protrusions on the Co islands and blue protrusions on the bare Cu(111) surface. Scan parameters: V = -1 mV, I=2 pA.

**Fig. 2:** **(A)** SP-dI/dV spectra of two Co islands with opposite spin orientation (dashed lines), as well as SP-dI/dV spectra of Fe adatoms on these two islands (solid lines). Geometry-induced fluctuations are minimized by averaging spectra at 10 different points for each island and averaging spectra of 5 different atoms (on the same island) for each adatom spectrum. **(B)** SP-dI/dV spectra of two Co islands with opposite spin orientation (dashed lines), as well as SP-dI/dV spectra of Cr adatoms on these two islands (solid lines). Slight differences in Co island spectra compared to (A) are due to the use of a different SP tip. **(C)** Normalized Fe dI/dV spectra (see text) from (A). **(D)** Normalized Cr dI/dV spectra (see text) from (B). **(E)** Difference between spin-up and spin-down spectra for Co islands (dashed black line), Fe adatoms (green line), and Cr adatoms (orange line) (data taken from (C) and (D), curves are normalized to one at their extrema). **(F)** Normalized dI/dV spectra measured with a non-SP-tip held over 8 different Co islands (black lines) as well as corresponding Fe adatoms (8 green lines) and Cr adatoms (8 orange lines) on these islands. Initial tunneling parameters for (A) and (C): V = -25 meV, I=3 pA, and for (B) and (D): V = -50 meV, I=2 pA. Fe and Cr adatom spectra in (A)-(D) and (F) multiplied by a constant factor of 3 for better clarity in plots.

**Fig. 3:** **(A)** SP dI/dV map of Fe and Cr adatoms on ↓ and ↑ Co islands on Cu(111). Scan parameters: V = -0.365 V, I=20 pA, T=4.8 K. **(B)** and **(C)** Zoom-ins of areas marked by dashed lines on ↓ and ↑ islands in (A). **(D)** and **(E)** linescans through the centers of Fe and Cr adatoms on ↓ and ↑ islands respectively (marked by dashed lines in (B) and (C)).

**Fig. 4:** Calculated binding energies of ferromagnetic and antiferromagnetic configurations for Fe and Cr adatoms on a 2-ML high Co film on Cu(111). Error bars indicate the energy difference between hcp and fcc adatom adsorption sites. Cartoons depict the lowest-energy magnetic coupling configuration for Fe and Cr adatoms on the Co film.

**Figures**

*Fig. 1*

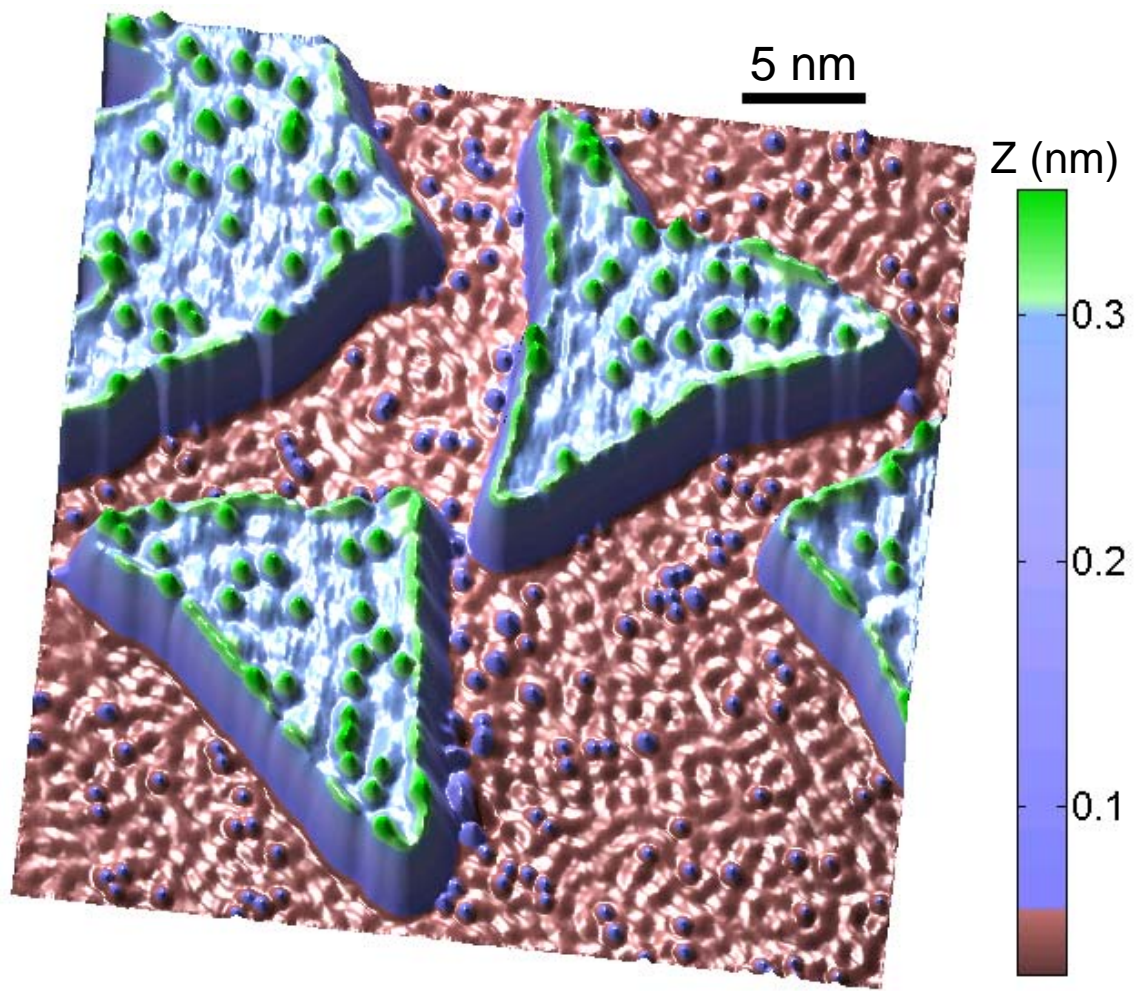

*Fig. 2*

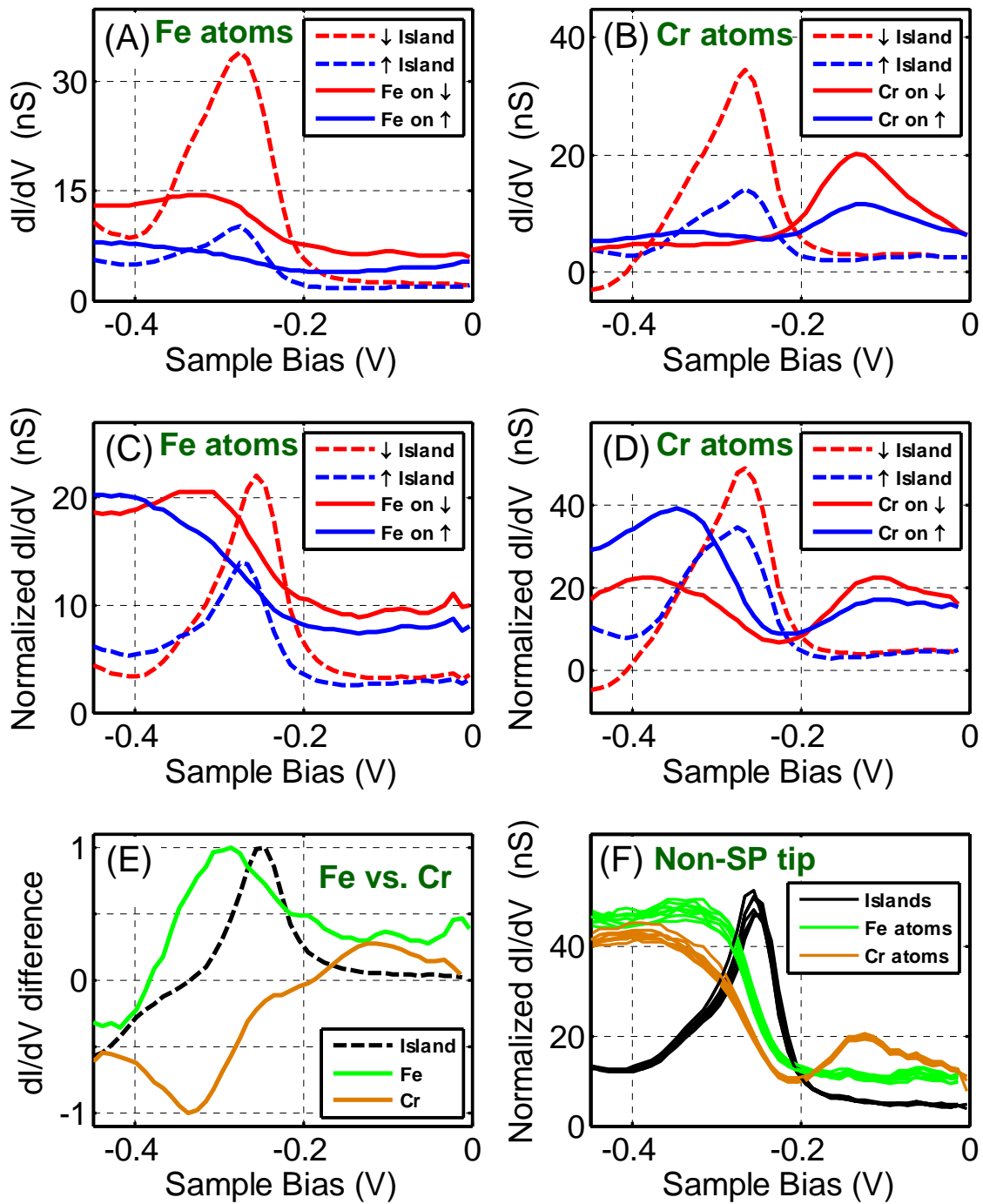

*Fig. 3*

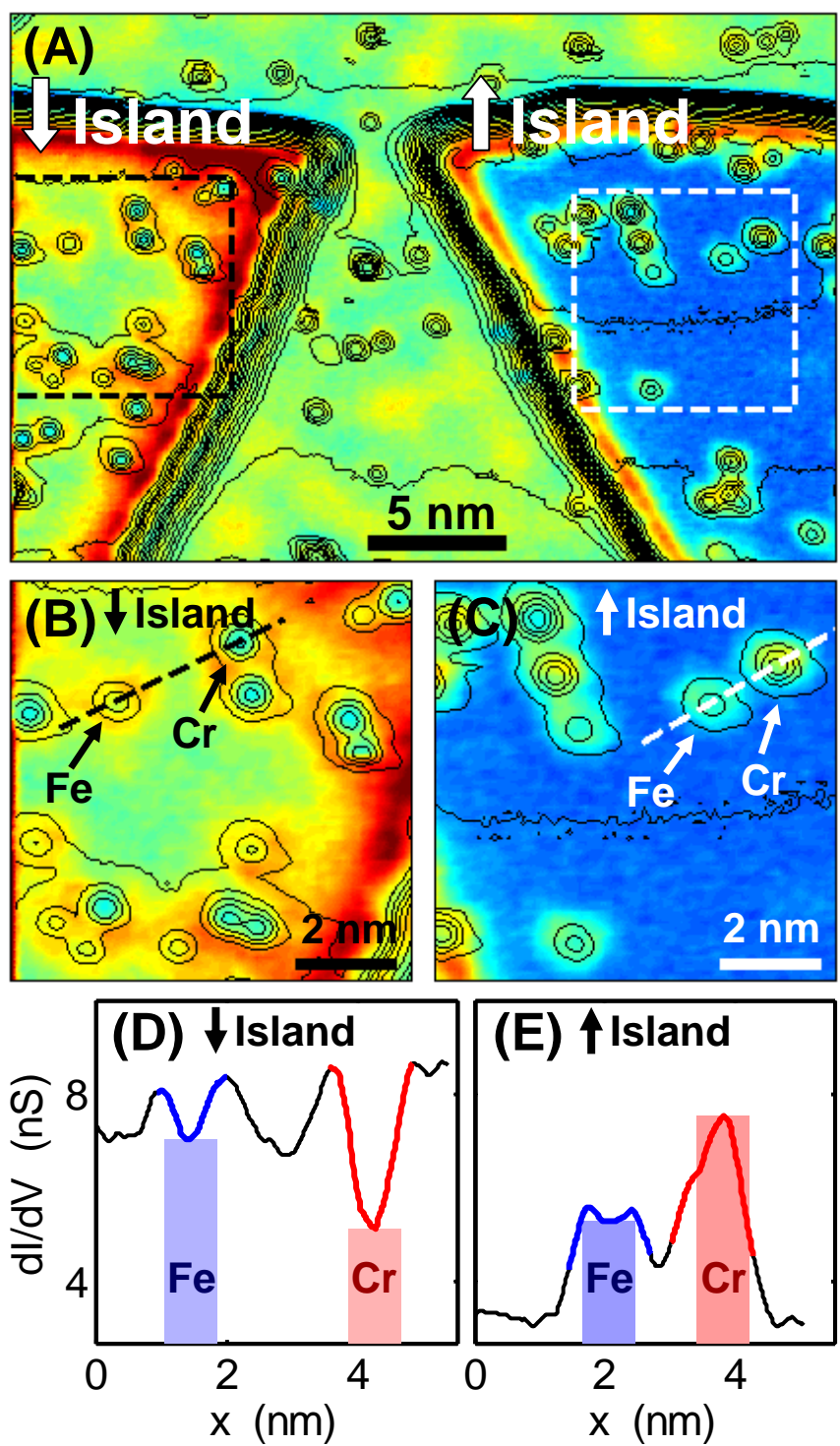

*Fig. 4*

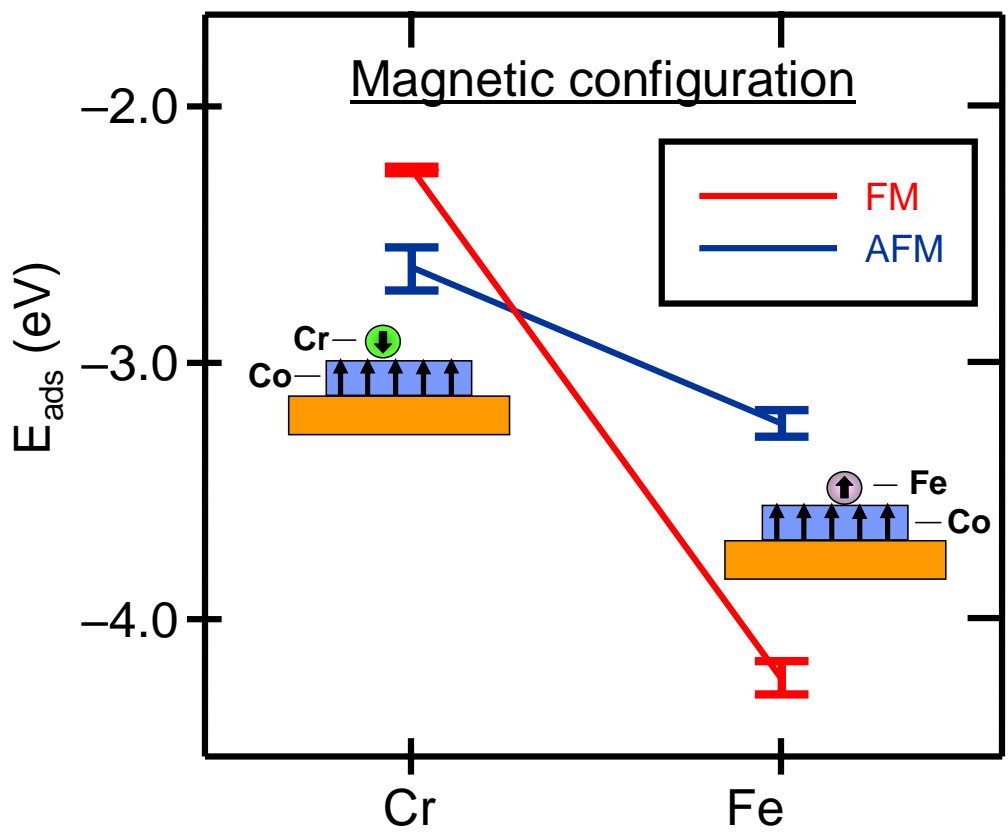